\documentclass[aps,prb,twocolumn,superscriptaddress,showpacs]{revtex4}
\usepackage{ae}
\usepackage[T1]{fontenc}
\usepackage[ansinew]{inputenc}
\usepackage{amsmath}
\usepackage{amssymb}
\usepackage{graphicx}
\usepackage{color}
\usepackage[colorlinks]{hyperref}
\usepackage{epstopdf}

\begin{document}

\date{\today}

\title{Scattering problems and boundary conditions
for 2D electron gas and  graphene. }

\author{A.M. Kadigrobov}
\affiliation{Theoretische Physik III, Ruhr-Universitaet Bochum, D-44801 Bochum, Germany }

%\date{\today}

\begin{abstract}
 Structure and coordinate dependence of the reflected  wave,  as well as boundary conditions
 for quasi-particles of graphene   and the two dimensional electron gas in sheets with  abrupt lattice edges
  are obtained and analyzed  by   the Green's function technique. In particular,  the reflection wave function
  contains terms inversely proportional to the distance to the graphene lattice edge. The Dirac equation and
  the momentum dependence
  of the wave functions of the quasi-particles   near the conical points are also found by the perturbation theory
  with degeneracy in  terms  of the Bloch functions taken at the degeneracy points. The developed approach allows
  to formulated the validity    criteria for the Dirac equation in a rather simple way.
\end{abstract}

\maketitle
\section{Introduction.}

Dynamical and  transport properties of various mesoscopic systems have been  attracting much attention
during the last decades \cite{Heinzel,Dittrich}. Among them are quantum dots, quantum nanowires, tunneling
junctions and 2D electron gas based nanostructures.Fascinating dynamic and kinetic  phenomena arise in  graphene
which is a two-dimensional (2D) semi-metal having  no energy gaps  between two  bands of electrons and holes at six
points of the hexagonal Brillouin zone.

 Electronic properties of graphene can be described by the two dimensional  differential
Dirac equation \cite{Wallace,Vincenzo} supplemented by boundary conditions. Details of the
boundary conditions depend on microscopic characteristics of the concrete structure of the
sample boundary\cite{Son}. Theoretical derivations of the boundary conditions for Dirac
equations are usually based on various models such as tight bound model (see, e.g., review
papers \cite{Neto,Sarma} and references there), the effective mass model \cite{Falko},
tight-binding model with a staggered potential at a zigzag boundary and with the boundary
orientation intermediate  between the zigzag and armchair forms \cite{AkhmerovBeenakker}.

In this paper dynamics of quasi-particles in 2D electron gas and graphene are considered in
the frame of the conventional approach to the   scattering problems for finite lattices in
terms of the electron Bloch functions and band energies without usage of the above-mentioned
models. Using the Green's function technique  the  boundary conditions and the coordinate
dependence of the wave functions of quasi-particles in 2D electron gas and graphene lattices
with an abrupt edges are obtained. Criteria of the validity of the Dirac equation are
formulated in a rather simple way. It is also shown that the wave function of the  reflected
quasi-particle contains slow varying terms inverse proportional to the distance to the edge of
the graphene sheet.

The outline of this paper is as follows.   In Sec.\ref{SectionDirac} the perturbation theory
with degeneracy is used to obtain the Dirac equation and the wave functions  in  terms of the
Bloch functions taken  at the degeneracy point in the reciprocal lattice. In
Sec.\ref{SectionPotential} scattering of quasi-particles by an external potential in graphene
is considered in the Bloch function representation.  The Dirac equation is derived and its
validity criteria are formulated. In Sec. IV   the Green's function approach to the problem of
scattering of quasi-particles in a lattice sheet with an abrupt edge is developed. In
Sec.\ref{Sec2Dboundary} the wave function structure and boundary conditions for the  2D
electron gas in a lattice sheet with an abrupt edge are found. In Sec.\ref{SecGraphBoundary},
the structure of the wave function and its dependence  on the distance to the lattice edge
 are found. In Sec.\ref{conclusion} concluding remarks are presented.

\section{Derivation of   Dirac equation and Bloch functions for graphene by perturbation theory. \label{SectionDirac}}
The Schr\"odinger equation for  electrons  is
\begin{eqnarray}
\hat{H}_0 \varphi_{s,\mathbf{p}}(\mathbf{r})=\varepsilon_s(\mathbf{p})
\varphi_{s,\mathbf{p}}(\mathbf{r})
 \label{Schroedinger0}
\end{eqnarray}
where $\hat{H}_0 $ is the Hamiltonian for electrons moving in the periodic lattice potential
$U(\mathbf{r+a})$ with the period $\mathbf{a}$. This Hamiltonian  reads as follows:
\begin{eqnarray}
\hat{H}_0 =-\frac{\hbar^2}{2 m}\frac{\partial^2}{\partial\mathbf{ r}^2} +U(\mathbf{r})
 \label{Hamiltonian0}
\end{eqnarray}
The wave function
\begin{eqnarray}
 \varphi_{s,\mathbf{p}}(\mathbf{r})=e^{i\frac{\mathbf{pr}}{\hbar}}u_{s,\mathbf{p}}(\mathbf{r})
 \label{Bloch}
\end{eqnarray}
is the Bloch function while the Bloch periodic factor $u_{s,\mathbf{p}}(\mathbf{r})$ has the
translation periodicity of the lattice, $\mathbf{p}$ is the electron quasi-momentum and $s$ is
the band number.

In order to find the dependence of Bloch functions and the dispersion law of the
quasi-particle in graphene on their  momentum $\mathbf{p}$ one may use the perturbation theory
in $|\mathbf{p}|/b\ll 1$ with the degeneracy\cite{LL,Slutskin} at $\mathbf{p}=0$ (here $b$ is
the characteristic period of the reciprocal lattice).

Presenting  Bloch functions  as a superposition  of the unperturbed wave functions of the zero
approximation
\begin{eqnarray}
\varphi(\mathbf{r})=\exp{i\frac{\mathbf{pr}}{\hbar}} \left(g_1 u_{1,0}(\mathbf{r}) + g_2
u_{,0}(\mathbf{r}) \right)
 \label{PerturbExpasion}
\end{eqnarray}
(here   $u_{1,2;0}(\mathbf{r})$ are the periodic factors  of the Bloch functions of the
degenerated bands taken at the point of degeneracy $\mathbf{p}=0$) and inserting it in the
Schr\"odinger equation Eq.(\ref{Schroedinger0}), after taking the matrix elements one gets a
set of algebraic equations for the expansion constants $g_{1,2}$. In the first approximation
in the momentum $\mathbf{p} $ these equations are:
\begin{eqnarray}
 -\bar{\varepsilon} g_{\alpha,\mathbf{p}}+ \sum_{\alpha{\prime}=1,2}
 \mathbf{p v_{\alpha,\alpha{\prime}}}g_{\alpha^{\prime},\mathbf{p}} =0
 \label{CoventEquation}
\end{eqnarray}
where $\alpha=1,2$ is the band number of the degenerated  band while $\bar{\varepsilon} \equiv
\varepsilon_{\alpha}(\mathbf{p})$;  the matrix elements of the velocity operator
$\hat{\mathbf{v}}=(-i\hbar/m)\partial/\partial\mathbf{ r} $  are
\begin{eqnarray}
\mathbf{v}_{\alpha,\alpha^{\prime}}= \int u_{\alpha,0}^{\star}(\mathbf{r})
\hat{\mathbf{v}}u_{\alpha^{\prime},0}(\mathbf{r})d\mathbf{r}
 \label{velocity}
\end{eqnarray}

Equating the determinant of  Eq.(\ref{CoventEquation}) one gets the conventional dispersion
law of quasi-particles near the degeneration point:
\begin{eqnarray}
\varepsilon_{\pm}(\mathbf{p})=\frac{\mathbf{pv_{+}}\pm
\sqrt{(\mathbf{pv_{-}})^2 + 4 |\mathbf{pv}_{12}|^2}}{2}
 \label{ConventionalDispersion}
\end{eqnarray}
where  $v_{\pm}
=\mathbf{v}_{11} \pm \mathbf{v}_{22}$.

From Eq.(\ref{ConventionalDispersion}) it follows that  the  dispersion  law of  quasi-particles in
the vicinity of  the band intersection is of the  graphene-type
\begin{eqnarray}
\varepsilon_{\pm}(\mathbf{p})=\pm v \sqrt{p_x^2 + p_y^2}= \pm vp
 \label{GrapheneDispersionLaw}
\end{eqnarray}
if the lattice symmetry imposes the following conditions  on the velocity matrix elements at
the point $p=0$ of the degeneration $\varepsilon_{1}(0)=\varepsilon_{2}(0)$:
\begin{eqnarray}
\mathbf{v}_{11}(0)=\mathbf{v}_{22}(0)=0, \hspace{0.2cm} |\mathbf{v}_{12}(0)| =v, \nonumber
\\
 v_{12}^{y}(0) =\pm i v_{12}^{(x)}(0) \hspace{1.5cm}
 \label{VelocityMatixElements}
\end{eqnarray}
where $v=v_F \approx 1 \times 10^6$ m/s for graphene.

Inserting these values in Eq.(\ref{CoventEquation})   one gets  equation for dependence of the
 the expansion coefficients on the momentum $\mathbf{p}$ as follows:
\begin{eqnarray}
-\bar{\varepsilon} g_{1,\mathbf{p}}
+ v(p_x -i p_y)g_{2,\mathbf{p}}=0; \nonumber \\
 v(p_x+i
p_y)g_{1,\mathbf{p}}-\bar{\varepsilon} g_{2,\mathbf{p}}  =0
 \label{Dirac1}
\end{eqnarray}
Using Eq.(\ref{Dirac1})  one finds the
dispersion law  Eq.(\ref{GrapheneDispersionLaw}) and the Bloch functions of
quasi-particles in graphene:
\begin{eqnarray}
\varphi_{\alpha,
\mathbf{p}}^{(gr;\pm)}(\mathbf{r})=g_1\exp{i\frac{\mathbf{pr}}{\hbar}}\nonumber \\
\times \left(%
\begin{array}{cc}
  u_{1,0}^{(\pm)}(\mathbf{r}) & 0 \\
 0 &  u_{2,0}^{(\pm)}(\mathbf{r})\\
\end{array}%
\right)\left(%
\begin{array}{c}
  1 \\
 e^{\mp i \theta} \\
\end{array}%
\right)
 \label{GraphBlochFunction}
\end{eqnarray}
where $g_1$ is the normalizing constant and the phase $\theta=\arctan(p_x/p_y)$.

In the next section, scattering of quasi-particles by an external potential  in 2D gas and
graphene is considered.

\section{Scattering of quasi-particles by an external potential and derivation of the Dirac equation. \label{SectionPotential}}
In this section,  the scattering problem of electrons by a potential $V(\mathbf{r})$ (the
characteristic properties of which are later described) in 2D gas and graphene is
investigated.

The Schr\"odinger equation of the system under consideration is
\begin{eqnarray}
\left(\hat{H}_0 +V(\mathbf{r})\right)\Psi(\mathbf{r})=\varepsilon \Psi(\mathbf{r})
 \label{SchroedingerV}
\end{eqnarray}
It is  assumed that   two energy bands are closely spaced or intersect in a certain point
$\mathbf{p}=0$ of the reciprocal space as it takes place in graphene while the energy
$\varepsilon$ is in  the vicinity of the degenerate energy. In order to investigate dynamics
of electrons in such a situation it is convenient  to expand  $\Psi$ in the series of
 the following   functions :
\begin{eqnarray}
\chi_{s,\mathbf{p}}=\exp{\left(i\frac{\mathbf{pr}}{\hbar}\right)}\left\{\begin{array}{c}
u_{\alpha,0}(\mathbf{r}), \hspace{0.2cm} s \equiv \alpha=1,2 \\
 u_{s,\mathbf{p}}(\mathbf{r}), \hspace{0.2cm} s \neq 1,2\\
\end{array} \right.
 \label{chi}
\end{eqnarray}
where band numbers $\alpha =1,2$ designate the bands close to each other, the periodic Bloch
factors of which are taken at $\mathbf{p}=0$. As $\chi_{s,\mathbf{p}}$ constitute a complete
set of functions the following expansion is satisfied for all values of $\mathbf{p}$.
\begin{eqnarray}
\Psi(\mathbf{r})=\sum_s \int g_{s,\mathbf{p}}\chi_{s,\mathbf{p}}(\mathbf{r})d\mathbf{p}
 \label{expansion}
\end{eqnarray}

Inserting this expansion in Eq.(\ref{SchroedingerV}) one gets
\begin{eqnarray}
\sum_{\alpha=1,2} \int d\mathbf{p}
g_{\alpha,\mathbf{p}}\exp{\left(i\frac{\mathbf{pr}}{\hbar}\right)}\Big(\varepsilon_\alpha
(0)+\mathbf{p\hat{v}}-\nonumber \\
+ \frac{\mathbf{p}^2}{2m} + V(\mathbf{r})-\varepsilon\Big)u_{\alpha,0}(\mathbf{r})
 \nonumber \\
 +\sum_{s
\neq 1,2} \int d\mathbf{p} g_{s,\mathbf{p}}\Big(\varepsilon_s
(\mathbf{p})+\mathbf{p\hat{v}}-\varepsilon\Big)\varphi_{s,\mathbf{p}}(\mathbf{r})=0
 \label{ExpansionEquation}
\end{eqnarray}
where  $\hat{\mathbf{v}}=-(i\hbar/m )\partial/\partial \mathbf{r} $ is the velocity operator.

Multiplying this equation on the left by $\chi_{\alpha,\mathbf{p}}$ and $\chi_{s,\mathbf{p}}$
in turns and integrating one gets a set of coupled algebraic equations for the expansion
factors $g_{\alpha,\mathbf{p}}$:
\begin{eqnarray}
\Big(\varepsilon_\alpha
(0)+\frac{\mathbf{p}^2}{2m}  -\varepsilon\Big)g_{\alpha,\mathbf{p}}+  \sum_{\alpha{\prime}=1,2}
 \mathbf{p v_{\alpha,\alpha{\prime}}}g_{\alpha^{\prime},\mathbf{p}}  \nonumber \\
+\int_{-\infty}^{\infty} V_{p-p^{\prime}}g_{\alpha,\mathbf{p}^{\prime}}d
\mathbf{p}^{\prime}\nonumber \\
 + \sum_{s \neq 1,2} \Big(\varepsilon_s (\mathbf{p})
-\varepsilon\Big)a_{\alpha,s}(\mathbf{p}) g_{s,\mathbf{p}} \nonumber
\\ +\sum_{s \neq 1,2} \int V_{p-p^{\prime}} a_{s,s^{\prime}}(\mathbf{p,p^{\prime}})
g_{s,\mathbf{p^{\prime}}}d\mathbf{p^{\prime}}=0 \hspace{0.3cm}
 \label{ExpansionEquation1}
\end{eqnarray}
\begin{eqnarray}
\Big(\varepsilon_s(\mathbf{p})-\varepsilon\Big)g_{s,\mathbf{p}}+
\sum_{\alpha=1,2}\Big\{\Big(\varepsilon_\alpha (0)+\frac{\mathbf{p}^2}{2m}  -\varepsilon\Big)
a^{\star}_{\alpha,s}g_{\alpha,\mathbf{p}}\nonumber \\
+\mathbf{p v_{s,\alpha}}g_{\alpha,\mathbf{p}} +\int_{-\infty}^{\infty}
V_{p-p^{\prime}}a^{\star}_{\alpha,s}(\mathbf{p^{\prime}}) g_{\alpha,\mathbf{p}^{\prime}}d
\mathbf{p}^{\prime}\Big\}\nonumber
\\
+\sum_{s \neq 1,2} \int V_{p-p^{\prime}} a_{s,s^{\prime}}(\mathbf{p,p^{\prime}})
g_{s^{\prime},\mathbf{p^{\prime}}}d\mathbf{p^{\prime}}=0\hspace{0.6cm}
 \label{ExpansionEquation2}
\end{eqnarray}
where the matrix elements of the velocity operator
$\hat{\mathbf{v}}=(-i\hbar/m)\partial/\partial\mathbf{ r} $ are
\begin{eqnarray}
\mathbf{v}_{\alpha,\alpha^{\prime}}= \oint_{(\mathbf{a})}u_{\alpha,0}^{\star}(\mathbf{r})
\hat{\mathbf{v}}u_{\alpha^{\prime},0}(\mathbf{r})d\mathbf{r } \nonumber \\
\mathbf{v}_{s,\alpha^{\prime}}=\oint_{(\mathbf{a})}u_{s,\mathbf{p}}^{\star}(\mathbf{r})\hat{\mathbf{v}}u_{\alpha^{\prime},0}(\mathbf{r})d\mathbf{r}
 \label{velocity1}
\end{eqnarray}
and
\begin{eqnarray}
a_{\alpha,s}(\mathbf{p})= \oint_{(\mathbf{a})}u_{\alpha,0}^{\star}(\mathbf{r})
u_{s,\mathbf{p}}(\mathbf{r})d\mathbf{r}  \nonumber \\
a_{s,s^{\prime}}(\mathbf{p,p}^{\prime})=\oint_{(\mathbf{a})}u_{s,\mathbf{p}}^{\star}(\mathbf{r})
u_{s^{\prime},\mathbf{p}^{\prime}}(\mathbf{r})d\mathbf{r}
 \label{a}
\end{eqnarray}
Integration in Eqs(\ref{velocity1},\ref{a}) is over a unit cell. The above equations  are valid
for all values of the electron momentum $\mathbf{p}$ and for an arbitrary form of the
potential $V(\mathbf{r})$.

The equations which describe dynamics of electrons in graphene and analogous conductors (Dirac
equations) in the vicinity of the degeneration energy are readily follows from
Eqs(\ref{ExpansionEquation1},\ref{ExpansionEquation2}) in the following limits: $|\mathbf{p}|
\ll b=2 \pi \hbar/a$ (where $a$ is the characteristic atomic spacing), the potential is small
and it slowly varies  that is $|V| \ll \Delta_{gap} = |\varepsilon_s(\mathbf{p}) -
\varepsilon_{\alpha}(\mathbf{p})|$, $s \neq \alpha$, and the characteristic interval $\delta
l$ of the variation of $V(\mathbf{r})$ is $\delta l \gg a$.

Indeed, under the above assumptions  one may neglect the dependence of the matrix elements  in
Eq.(\ref{a}) on $\mathbf{p}$ and obtain  $a_{\alpha,s}=\delta_{\alpha,s} =0, \; s\neq \alpha$.
Inserting this equality in Eq.(\ref{ExpansionEquation2})   one gets
\begin{eqnarray}
\gamma =\frac{|g_{s,\mathbf{p}}|}{|g_{\alpha,\mathbf{p}}|} \sim \frac{|V|,
(p^2/2m)}{\Delta_{gap}}\ll 1
 \label{Inequality}
\end{eqnarray}
 and hence  equations Eq.(\ref{ExpansionEquation1})
and Eq,(\ref{ExpansionEquation2}) are decoupled in  the zero approximation in $\gamma \ll 1$.
Therefore, in this approximation the Schr\"odinger equation in the $\chi$-representation (see
Eq.(\ref{expansion})) for electrons   in the vicinity of the intersection of two bands,
$\varepsilon_1(0)=\varepsilon_2(0)=0$, reads as follows:
\begin{eqnarray}
\Big(\varepsilon_\alpha (0) -\varepsilon\Big)g_{\alpha,\mathbf{p}}+ \sum_{\alpha{\prime}=1,2}
 \mathbf{p v_{\alpha,\alpha{\prime}}}g_{\alpha^{\prime},\mathbf{p}}\nonumber \\
 +
 \int_{-\infty}^{\infty} V_{p-p^{\prime}}g_{\alpha,\mathbf{p}^{\prime}}d \mathbf{p}^{\prime}=0
 \label{ReducedHammiltonian}
\end{eqnarray}
While writing this equation   we  assumed $p^2/2 m \ll |\mathbf{vp}|$).

Using equalities Eq.(\ref{VelocityMatixElements}) one gets the set of  equations that
describes dynamics of quasi-particles in the presence of  potential $V(x)$:
\begin{eqnarray}
 -\varepsilon g_{1,\mathbf{p}}+\int_{-\infty}^{\infty}
V_{p-p^{\prime}}g_{1,\mathbf{p}^{\prime}}d \mathbf{p}^{\prime}
+ v(p_x -i p_y)g_{2,\mathbf{p}}=0; \nonumber \\
 v(p_x+i
p_y)g_{1,\mathbf{p}} -\varepsilon g_{2,\mathbf{p}}+\int_{-\infty}^{\infty}
V_{p-p^{\prime}}g_{2,\mathbf{p}^{\prime}}d \mathbf{p}^{\prime} =0 \,\,
 \label{pEquation}
\end{eqnarray}
where for the sake of certainty $v_{12}^{y} =-i v_{12}^{(x)} $ is chosen. Expanding the wave
functions in Eq.(\ref{pEquation}) into the Fourier series
\begin{eqnarray}
g_{1,2;\mathbf{p}}=\int \Phi_{1,2}(\mathbf{r})\exp\{-i \frac{p \mathbf{r}}{\hbar}\}d
\mathbf{r}
 \label{Dirac}
\end{eqnarray}
one find the equation for the Fourier factors:
\begin{eqnarray}
-i \hbar v\left(%
\begin{array}{cc}
  V(\mathbf{r}) & \partial_x -i \partial_y \\
\partial_x+i \partial_y  &V(\mathbf{r})\\
\end{array}%
\right)\left(%
\begin{array}{c}
  \Phi_1 \\
  \Phi_2\\
\end{array}%
\right) =\varepsilon \left(%
\begin{array}{c}
  \Phi_1 \\
  \Phi_2\\
\end{array}%
\right)
 \label{DiracEquation}
\end{eqnarray}
The above equation describes dynamics and, in particular,  quantum tunnelling of
quasi-particles between intersecting   energy bands in  the vicinity of the point of
degeneration. This set of differential equations (with proper changes of parameters) arises in
all cases in which the unperturbed energy spectrum has points of degeneration (or points of
close approach of energy bands), e.g., in graphene (see review papers
 \cite{Beenakker,Neto,Sarma}), in the cases of  Landau-Zener tunnelling (see Ref.\cite{LLZener} and references
 there) and the magnetic breakdown - quantum tunnelling in metals under a strong magnetic
 field (see Refs.\cite{Blount,Slutskin,SK}). Note, that the tunnelling transmission of electrons
  between intersecting energy bands  without back-scattering ("Klein tunnelling") takes place in many  cases,
  in particular, in the cases of
  grapene\cite{Beenakker,Neto,Sarma}  (normal incident of the electron to
  barrier) and the magnetic breakdown \cite{Slutskin}.

As it follows from  the above derivation of Eq.(\ref{DiracEquation}) the Dirac
equation\cite{Beenakker,Neto} is valid only in the limit of small and smooth potentials (see
Eq.(\ref{Inequality}) and the text around it) and hence it can not be used for investigation
of the problem of electron scattering  by sharp boundaries of the sample. In the next section
the Green function approach is developed to solve  this problem for the cases of 2D gas and
graphene.

\section{Scattering of electros in lattice with  abrupt edge  (Green function approach). \label{GeneralEdge}}

Boundary conditions for Dirac Fermions in graphene were derived in References
\cite{Falko,AkhmerovBeenakker,Beenakker}(see also, e.g.,  Review papers \cite{Neto,Sarma}) in
the tight-binding model. In this section the boundary conditions for two dimensional electron
gas and grapene are derived by use of the Green's function technique  in
 terms of the general properties of   electron spectra and proper wave functions.

Below  a half infinite two dimensional sheet  of 2D gas or graphene  in the half plane $x \geq
0$ with the edge line at $x=0$ is considered. In this case the Schr\"odinger equation is
\begin{eqnarray}
\left(-\frac{\hbar^2}{2 m}\frac{\partial^2}{\partial\mathbf{ r}^2} +U(\mathbf{r})
-\varepsilon\right)\Psi(\mathbf{r})=0
 \label{SchroedingerBoundary}
\end{eqnarray}
where  $U(\mathbf{r})$ is the periodic lattice potential. For the sake of certainty  the
boundary conditions
\begin{eqnarray}
\Psi(\mathbf{r})&=&0, \hspace{1.4cm} x=0  \nonumber \\
\Psi (\mathbf{r}) &=& \varphi_{s_0,\mathbf{p}_0}^{(in)}(\mathbf{r}) , \;\; x\rightarrow
+\infty
 \label{BoundaryCondition1}
\end{eqnarray}
are   assumed. Here $\varphi_{s_0,\mathbf{p}_0}^{(in)}$ is the Bloch function (see
Eq.(\ref{Bloch})) incident to the boundary and $\mathbf{p}_0=(p_x^{(0)}, q)$ where
$q=p_y^{(in)}$ is the conserving momentum projection.

Below, to investigate the  problem of scattering by the abrupt edge at $x=0$ Green's function
for the infinite lattice is used, that is the needed  Green function satisfies
 the equation
\begin{eqnarray}
\left(-\frac{\hbar^2}{2 m}\frac{\partial^2}{\partial\mathbf{ r}^2} +U(\mathbf{r})
-\varepsilon\right)G(\mathbf{r},\mathbf{r}^{\prime})=\delta(\mathbf{r}-\mathbf{r}^{\prime})
 \label{GreenFunctionEq}
\end{eqnarray}
in which the  lattice potential $U(\mathbf{r})$ covers the whole plane $(x,y)$. Expanding
$G(\mathbf{r},\mathbf{r}^{\prime})$  in the series of wave functions of electrons in the
infinite lattice  and using Eq.(\ref{GreenFunctionEq}) one  finds
\begin{eqnarray}
G(\mathbf{r},\mathbf{r}^{\prime})=\sum_s\int\frac{\varphi_{s,\mathbf{p}}^{\star}(\mathbf{r})
\varphi_{s,\mathbf{p}}(\mathbf{r}^{\prime})}{\varepsilon-\varepsilon_s(\mathbf{p})+i \delta}d
\mathbf{p}
 \label{GreenFunctionExpanded}
\end{eqnarray}
where $\delta \rightarrow +0$

\begin{figure}
\centerline{\includegraphics[width=8.0cm]{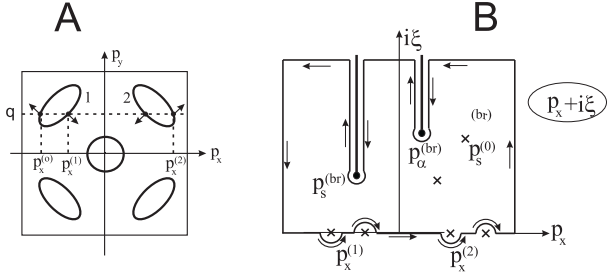}} \caption{A. An example of the
Brillouin zone with 5 contours of equal energy  of energy bands
$\varepsilon_{\alpha}(\mathbf{p}) =\varepsilon $  with the band number $\alpha=1,2,3,4,5$. The
arrows show the directions of the velocities at the points of intersections of the equal
energy contours with the line of the constant $y$-projection $q$.  The incident electron has
the conserving momentum projection $q=p_y^{(in)}$ and the negative direction of the velocity
$x$-projection, $v_x$.  B. Contour of integration of
Eq.(\ref{FunctionPsi},\ref{ComplexIntegral}). Crosses on the real axis  $p_x$ and those in the
upper complex half-plane   show positions of the poles corresponding to two points with
positive $v_x$ and to   virtual  states $\varepsilon_{s}(\mathbf{p}) \neq \varepsilon; \; s
\neq \alpha $. Thick vertical lines are brunch lines corresponding to the brunching points
(thick dots) in the electron spectrum.\label{FigureContour}}
\end{figure}
%%%%%%%%%%%%%%%%%%%%%%%%%%%%

Below we also assume that along the edge line $x=0$   the lattice is periodic with the period
$\bar{a}$ and hence the momentum projection $p_y$ conserves. Taking into account this
requirement    and using Eqs.(\ref{SchroedingerBoundary},\ref{GreenFunctionEq})  together with
Eq.(\ref{GreenFunctionExpanded}) and  the boundary conditions for $\Psi(\mathbf{r})$ one finds
the wave function  on the right half-plane $x \geq 0$ as follows:
\begin{eqnarray}
\Psi(\mathbf{r})=   \frac{\varphi_{s_0,\mathbf{p}_0}^{(in)}}{\sqrt{v_{x,s_0}}}  +
\frac{\hbar^2}{2 m}\int_{-\infty}^{\infty} d\bar{y}
\Psi_{x}^{\prime}(-0,\bar{y}) \nonumber \\
 \times \sum_s \int dp_x\frac{\varphi_{s,p_x,q}^{\star}(-0,\bar{y})
\varphi_{s,p_x,q_y}(\mathbf{r})}{\varepsilon-\varepsilon_s(p_x,q_y)+i \delta}
 \label{FunctionPsi}
\end{eqnarray}
where  $v_{x,s_0} < 0$ is the $x$-projection of the velocity of the incident electron that
normalizes its wave function to the unity flux; $q_y \equiv p_0^{y}$ is the conserving
$y$-projection of the momentum $\mathbf{p}_0$ of  the incident electron;
$\Psi_x^{\prime}(-0,\bar{y})=\partial\Psi(\mathbf{r})/\partial x$ at $x=-0$; as the value of
$\Psi$-function in Eq.(\ref{FunctionPsi}) exactly on the boundary contour is a matter of
convention (see Ref.\cite{Morse}) the boundary contour is assumed to be shifted to $x=-0
\equiv 0-\bar{\delta}, \;\bar{\delta} \rightarrow 0$ while $\Psi(\mathbf{r})$ is defined on
the half-plane $x \geq 0$.

It is now necessary to introduce the integral equation for $ \Psi_{x}^{\prime}(-0,\bar{y})$
solution of which completes the definition of the wave function $\Psi(\mathbf{r})$:
\begin{eqnarray}
\Psi_{x}^{\prime}(0,y)= \frac{1}{\sqrt{v_{x,s_0}}}  \frac{\partial
\varphi_{s_0,\mathbf{p}_0}^{(in)}}{\partial x}\Big|_{x=0} \nonumber \\
  +\frac{\hbar^2}{2m}
\int_{-\infty}^{\infty} d\bar{y} \Psi_{x}^{\prime}(0,\bar{y}) \nonumber \\
 \times \sum_s \int
dp_x\frac{\varphi_{s,p_x,q}^{\star}(0,\bar{y})
\varphi^{\prime}_{s,p_x,q_y}(0,y)}{\varepsilon-\varepsilon_s(p_x,q)+i \delta}\hspace{0.2cm}
 \label{PsiDerivativeEquation}
\end{eqnarray}
Here $f^{\prime}(\mathbf{r})= \partial f(\mathbf{r})/\partial x$.

In the general case and without usage of an approximate model this integral equation can not
be solved. However,
 important general features (in terms of  $\Psi_{x}^{\prime}(0,\bar{y})$)  of the quasi-particle scattering    by the
abrupt lattice edge   follow from  Eq.(\ref{FunctionPsi}).

Indeed, let us consider  one-dimensional integrals in Eq.(\ref{FunctionPsi}) presenting them
in the form
\begin{eqnarray}
I=\int_{-b_x/2}^{b_x/2} \frac{u_{s,p_x,q_y}^{\star}(0,\bar{y})
u_{s,p_x,q_y}(\mathbf{r})e^{i x p_x/\hbar}}{\varepsilon-\varepsilon_s(p_x,q_y)+i \delta}dp_x
 \label{ComplexIntegral}
\end{eqnarray}
%%%%%%%%%%%%%%%%%%%%%%%%%%
Here $b_x$ is the period of the reciprocal lattice in the $x$-direction. In the complex plane
the dispersion law $\varepsilon_s(p_x,q)$  considered as a function of the complex variable
$z=p_x+i\xi$   is a multi-valued function which has branching points in the complex plane and
hence this integral is a sum of the residues minus sum of integrals along  the brunch cuts in
the upper complex half-plane $\xi \geq 0$ inside the contour schematically shown in
Fig.\ref{FigureContour}. The left and right vertical lines of the contour are separated by the
reciprocal period period $b_x$ and hence the integrals along them cancel each other because
the integrands are periodic functions of the same  period. The integral along its upper
horizonal part exponentially goes to zero as this contour part goes to $i \infty$.

The poles and branching cuts  of the integrand which contribute to the integral
Eq.(\ref{ComplexIntegral}) are separated in two types:

1. Poles lying on the upper side of the real axis $$p_x=p_x^{(\alpha)}
+i\frac{\delta}{v_x^{(\alpha)}}, \; \alpha=1,2,..., \; \delta \rightarrow 0$$ where
their real parts $p_x^{(\alpha)}$ are determined by the equation
\begin{eqnarray}
\varepsilon=\varepsilon_\alpha(p_x,q_y)
\label{EqualEnergy}
\end{eqnarray}
%%%%%%%%%%%%%%%%%%%%%%%%%%
while $\alpha$ defines the number of  the band which are present in the infinite lattice at
the energy $\varepsilon$  and the momentum projection $p_y=q$ (in Fig.\ref{FigureContour} they
are shown as pockets with the band numbers $\alpha= 1,2$). One easily sees that these poles
are inside the integration contour if the x-projections of the velocity
\begin{eqnarray}
v_{x,\alpha}=\frac{\partial  \varepsilon_\alpha(p_x,q_y)}{\partial p_x
}\Big|_{p_x^{(\alpha)}}>0 \label{reflectionvelocty}
\end{eqnarray}
and, therefore, they correspond to the states of electrons reflected back by the boundary.

2. Poles   lying high  in the upper complex plane $p_s^{(0)} =p_x^{s} +i b_s^{(0)} $ which are
determined by the equation $\varepsilon=\varepsilon_s(p_x,q_y), \; s \neq \alpha$ where the
energy bands $\varepsilon_s(p_x,q_y)$ do not overlap
 bands $\alpha$ (in which the energy $\varepsilon$ lies).

 As the dispersion law $\varepsilon_s(p_x,q)$ is a multi-valued  functions of $p_x$
 (a circuit around the branching point  changes the band number $s$)
there are
 branching cuts in the upper half plane  $z=p_x +i \xi$, schematically shown in
 Fig.\ref{FigureContour}, which pass  from the branching points
 $p_s^{br)}=p_{x,s}^{(br)} + i \xi_s^{(br)}$ to  $p_{x,s}^{(br)}+i \infty$).

Taking into account the above-mentioned poles and branch cuts one easily  carried out
integration  in  Eq.(\ref{ComplexIntegral}) and,  inserting the result in
Eq.(\ref{FunctionPsi}), one writes the required wave function as follows:
\begin{eqnarray}
\Psi(\mathbf{r})=  \frac{\varphi_{s_0,\mathbf{p}_0}^{(in)}}{\sqrt{v_{x,s_0}}} +
\sum_{\alpha}C_{\alpha}
\frac{ \varphi_{\alpha,p_x^{\alpha}}(x,y)}{\sqrt{v_{x,\alpha}}} \nonumber \\
+\left[\sum_{s \neq \alpha}  C_s\frac{ \phi_{s, p_s^{(0)}}(x,y)}{\sqrt{v_{x,s}}}
 e^{-b_s^{(0)} x/\hbar}\right. \hspace{1.5cm} \nonumber \\
+\left. \sum_{\bar{s}} \int_{\xi_{br}^{(\bar{s})}}^{\infty}
B_{\bar{s}}(\zeta)\phi_{\bar{s},p_{\bar{s}}^{(br)}(\zeta)}(x,y) e^{-x\zeta/\hbar} d \zeta
\right]
 \label{PsiFinal}
\end{eqnarray}
where summation $\bar{s}=s,\alpha$ (that is summation with respect to $\bar{s}$ is over all
electron bands including all $\alpha$-bands), the integral is taken along the
$\bar{s}$-branching cut in which the variable change $\zeta=i \xi$ has been done; the
functions in square brackets are
\begin{eqnarray}
\phi_{s,p_s^{(0)}}=u_{s,p_s^{(0)}}\exp\{i\frac { x p_x^{(s)}+ qy}{\hbar}\} \hspace{1.2cm} \nonumber \\
\phi_{\bar{s},p_{\bar{s}}^{(br)}(\zeta)}=u_{\bar{s},p_{\bar{s}}^{(br)}(\zeta)}\exp\{i\frac { x
p_{x,\bar{s}}^{(br)})+ qy}{\hbar}\}
 \label{phifunctions}
\end{eqnarray}
 where $p_s^{(br)}(\zeta)=p_{x,s}^{(br)}+\zeta$; constants $A_s, C_\alpha$ and
 $B_{\bar{s}}(\zeta)$ are presented in Appendix, Eqs.(\ref{amplitudes},\ref{B}).

\subsection{Scattering of electrons by abrupt edge in 2D electron gas. \label{Sec2Dboundary}}

As one sees from Eq.(\ref{PsiFinal}) the functions in square brackets exponentially decrease
with an increase of the x-coordinate. In the general case the energy gaps between
non-overlapping electron bands are of the order of the band widths $\Delta E \sim 1 \div 10$eV
and hence the imaginary parts of the pole and branch momenta
 are of the order of the $\Delta E/v_F$  that is  $b_s^{(0)} \sim \xi_{br}^{(\bar{s})} \sim \hbar/a $
 where $a$ is the atomic spacing.

 From the above considerations  and Eq.(\ref{PsiFinal}) it follows that   inside the layer
 $x \lesssim a$  adjacent to  the sample boundary the electron wave function is a
 superposition of  Bloch wave functions $\varphi_{\mathbf{s}}$  of all energy bands including those
virtual which are above and below the band of the incident electron
 (that is their band numbers
$s\neq \alpha$).

At the distances much larger than the atomic spacing, $x \gg a$, all the virtual wave
functions   exponentially  drop out from the superposition and the electro wave function
$\Psi$ reduces to
\begin{eqnarray}
\Psi(\mathbf{r})=   \varphi_{s_0,\mathbf{p}_0}^{(in)}  + \sum_{\alpha}C_{\alpha} \frac{
\varphi_{\alpha,p_x^{(\alpha)}}(x,y)}{\sqrt{v_{x,\alpha}}} +\textsl{O}(e^{-x/a})
 \label{PsiReduced}
\end{eqnarray}
According to the calculations the Bloch functions under the summation  sign belong to the
states with the positive x-projections of the electron velocity (see
Eq.(\ref{reflectionvelocty})  Fig.\ref{FigureContour}A). Therefore, they are the Bloch
functions of the electron scattered back by the sample boundary into all the available energy
bands at the energy of the incident electron $\varepsilon$ and the conserving $y$-projection
of the momentum  while constants $C_{\alpha}$  are the  probability amplitudes  of this
many-channel scattering (an example of such the two-channel scattering is presented in in
Fig.\ref{FigureContour}A).

The above general scattering scenario requires a special  treatment in the case of the
generation  when  the top and the bottom of two energy bands are very close or coincide in
some point of the reciprocal space as it takes place in graphene. In the next section the
scattering of quasi-particles by a sharp graphene boundary is considered.
\subsection{Scattering of quasi-particles by abrupt edge of graphene sheet.
\label{SecGraphBoundary}}

In this section  scattering   of  quasi-particles in  graphene by an abrupt edge is
considered. Graphene fills the half-plane $x\geq 0$ while the boundary condition for the
quasi-particle wave function is $\Psi^{(gr)}(0,y)=0$

\begin{figure}
\centerline{\includegraphics[width=8.0cm]{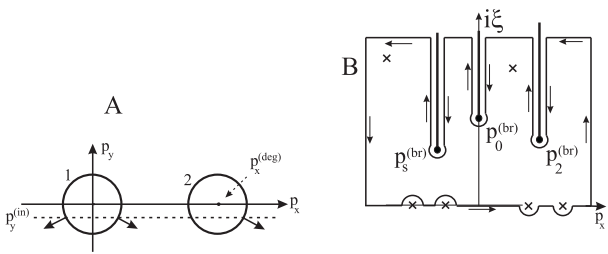}} \caption{A. Shematic
representation of two   equal energy  contours  $v\sqrt{p_x^2 +p_y^2} =\varepsilon $  and
$v\sqrt{(p_x-p_x^{deg})^2 +p_y^2} =\varepsilon $. The arrows show  directions of the
velocities at the points of intersections of the equal energy contours with the line of the
constant $y$-projection $q$. The incident electron has the conserving momentum projection
$q=p_y^{(in)}$ and the negative direction of the velocity $x$-projection, $v_x =v p_x/p$. B.
Contour of integration of Eq.(\ref{FunctionPsi},\ref{ComplexIntegral}). Crosses on the real
axis $p_x$ and those in the upper complex half-plane   show positions of the poles
corresponding to two points with positive $v_x$ and to   virtual  states
$\varepsilon_{s}(\mathbf{p}) \neq \varepsilon; \; s \neq \alpha $. Thick vertical lines are
brunch lines corresponding to the brunching points (thick dots) in the quasi-psrticle
spectrum.} \label{GraphContour}
\end{figure}
%%%%%%%%%%%%%%%%%%%%%%%%%%%%

In  the general  approach to the scattering problem developed above,  the only peculiarity of
the scattering of quasi-particles in graphene lies in their dispersion law whereas all the
equations of the previous section remain valid.

The  incident electron in graphene with a  negative $x$-projection of the velocity  in the
Bloch state $\varphi_{ \mathbf{p}^{(in)}}^{(gr,in)}$ and the energy $\varepsilon > 0$
belonging (for the sake ofcertainty)  to the electronic band $\varepsilon_{+}(p)=+v p$ is
considered (see Fig.\ref{GraphContour}A).

%For the sake of simplicity it is assumed that the line $p_y=p_{y}^{(0)}$ in the Brillouin zone
%crosses only one equal energy cone $\varepsilon(p)=v p$ (see Fig.\ref{FigGrapf}).

Inserting the graphene quasi-particle dispersion law   Eq.(\ref{GrapheneDispersionLaw}) in
Eq.(\ref{PsiFinal}) one finds the  electron wave function at the distances much greater that
the deBroglie's wave length $x \gg \lambda_B= \hbar v/\varepsilon$ as follows (details of
calculations are given in Appendix \ref{Appcutintegral}):
\begin{eqnarray}
x \gg \frac{\hbar v}{\varepsilon}, \hspace{6cm} \nonumber \\
 \Psi^{(gr)}(\mathbf{r})=
\frac{\varphi_{\mathbf{p}^{(in)}}^{(gr,in)}}{\sqrt{v_{x}}} +\sum_{\alpha=1,2}C_{\alpha}^{(gr)}
\frac{
\varphi_{\alpha,p_x^{\alpha},p_y^{(in)}}^{(gr;+)}(x,y)}{\sqrt{v_{x,\alpha}}}\nonumber \\
+\sum_{\nu=\pm}\sum_{\alpha=1,2}B_{\alpha}^{(gr;\nu)}\varphi_{\alpha;{p_x}^{(gr)} ,p_y^{(in)}}^{(gr;\nu)}(x,y)
\frac{e^{-x p_y^{(in)}/\hbar}}{x}
 \label{GraphPsiFinal}
\end{eqnarray}
Here  Bloch functions $\varphi_{\alpha,\mathbf{p}}^{(gr)}$ (see Eq.(11)  are
\begin{eqnarray}
\varphi_{\alpha,
\mathbf{p}}^{(gr;\pm)}(\mathbf{r})=g_{\alpha}\exp{i\frac{\mathbf{pr}}{\hbar}}\nonumber \\
 \left(%
\begin{array}{cc}
  u_{1,\mathbf{{p}}_{\alpha}^{(gr)} }^{(\pm)}(\mathbf{r}) & 0 \\
 0 &  u_{2,\mathbf{{p}}_{\alpha}^{(gr)} }^{(\pm)}(\mathbf{r})\\
\end{array}%
\right)\left(%
\begin{array}{c}
  1 \\
 e^{\mp i \theta} \\
\end{array}%
\right)
 \label{GraphBlochFunction1}
\end{eqnarray}
where  $g_{\alpha}$ is the normalizing constant    and    $\mathbf{{p}}_{1}^{(gr)} =0$ for
$\alpha =1$, while for $\alpha =2$ it is equal to the coordinate of the second cone $
\mathbf{p}^{(deg)}_2$; note that $p_y^{(in)}$ is the conserving $y$-projection of the incident
quasi-particle momentum.

Therefore, as it follows from Eqs.(\ref{PsiFinal},\ref{GraphPsiFinal}), near the graphene
lattice boundary, inside the layer  $x \lesssim a$ ($a$ is the atomic spacing), the
quasi-particle wave function is a superposition of Bloch wave functions belonging to all
energy bands (including those virtual). At the distance much larger than deBrouglie's wave
length $\lambda_B =\hbar v/\varepsilon$ the superposition reduces to the sum of the Bloch
functions of the reflected   electron,  Eq.(\ref{GraphBlochFunction}), (note, it was assumed
that an electron was the incident quasi-particle)
 of the infinite graphene sample  plus
additional terms   proportional to the graphene Bloch functions with  the momentum the both
projections of which are equal to the conserving $y$-projection of the incident quasi-particle
$p_y^{(in)}$. The latter terms   slowly decrease at the distances $\delta x \lesssim \hbar/|
p_y^{(in)}| \ll \varepsilon/v$. If the normal incidence of the quasi-article  on the graphene
boundary takes place, $p_y^{(in)}=0$, these terms decease as $1/x$.

\section{Conclusion. \label{conclusion}}

In this paper the problem of scattering of quasi-particles by an abrupt edge in the 2D
electron gas and in graphene lattice is considered by the Qreen's function technique. This
approach allows to find the  coordinate dependence of the wave function of the quasi-particle
reflected at such an  edge  and the boundary conditions in a rather simple way. In particular,
it is shown that the wave function of the reflected quasi-particle in graphene contains terms
slowly decreasing with an increase of the distance to the edge. In the case of the transverse
incidence they  are inverse proportional to this distance.

For graphene  the Dirac equation, the momentum dependence of the wave functions
 near the conic points and the dispersion law are derived  by the perturbation method with
degeneracy in terms of the Bloch functions the periodic  factors of which are taken at the
degeneracy point (the conic point).
 This approach allows to   formulate the  lattice symmetry
 and external field  properties   needed for validity of
the Dirac equation in  grahene and other two-dimensional conductors with degenerated energy
bands.

\textbf{\emph{Acknowledgement}}. The author  thanks   A.F. Volkov for useful discussions.

\begin{appendix}

\
\section{ \label{Constants}}

After taking \textbf{thew} integral  in Eq.(\ref{ComplexIntegral}) and the use of
Eq.(\ref{FunctionPsi}) one finds constants $C_{\alpha}, \; C_s$ and  function
$B_{\bar{s}}(\zeta)$ as follows:

\begin{eqnarray}
C_\alpha=2\pi i \frac{\hbar^2}{2 m}\int_{-\infty}^{\infty} d\bar{y}
\Psi_{x}^{\prime}(0,\bar{y}) \frac{\varphi^{\star}_{s, p_x^{(\alpha)}}(0,\bar{y})}{\sqrt{v_{x,\alpha}}} \nonumber \\
C_s=2\pi i \frac{\hbar^2}{2 m}\int_{-\infty}^{\infty} d\bar{y} \Psi_{x}^{\prime}(0,\bar{y})
\frac{\phi^{\star}_{s, p_s^{(0)}}(0,\bar{y})}{\sqrt{v_{x,s}}} \label{amplitudes}
\end{eqnarray}
and
\begin{eqnarray}
B_{\bar{s}}(\zeta)=\frac{  \hbar^2}{2 m}\int_{-\infty}^{\infty}  \frac{
\Psi_{x}^{\prime}(0,\bar{y}) \phi^{\star}_{\bar{s},
p_{\bar{s}}^{(br)}(\zeta)}(0,\bar{y})}{\varepsilon - \varepsilon_{\bar{s}}(p_{x,s}^{(br)}+
\zeta,q)) +i \delta} d\bar{y}.
 \label{B}
\end{eqnarray}
\section{Coordinate dependence of the integral along the cut for graphene. \label{Appcutintegral}}

Using the grapene dispersion law  Eq.(\ref{GrapheneDispersionLaw}) and Eq.(\ref{B}) one
re-writes the terms in  the last sum in Eq.(\ref{PsiFinal}) related to the energy bands
 with the grapene dispersion laws, $\alpha =1,2$,  as follows:
\begin{eqnarray}
\frac{\hbar^2}{2m}\int_{-\infty}^{\infty}d\bar{y}\Psi^{\prime}_x(0,\bar{y})
e^{ip_y^{(in)}(\bar{y}-y)/\hbar} \nonumber \\
 \times u_{\alpha;\mathbf{{p}}^{(gr)}}^{(gr)\star}(0,\bar{y})
u^{(gr)}_{\alpha;\mathbf{{p}}^{(gr)}}(\mathbf{r}) J_{\alpha}^{(gr)}(\bar{y},\mathbf{r})
 \label{graphB}
\end{eqnarray}
where
\begin{eqnarray}
J_{\alpha}^{(gr)}=\int_{q}^{\infty} \frac{ e^{-x \zeta/ \hbar}}{\varepsilon \mp v
\sqrt{-\zeta^2 +q^2}+i\delta}d\zeta
 \label{AppgraphIntegral1}
\end{eqnarray}
%Here $u_{\alpha,\bar{\mathbf{p}^{(gr)}}}^{(gr)}(\mathbf{r})$ are the periodic factors of the
%Bloch functions in graphene (see Eq.(\ref{garphBlochfunctions})) and $\mathbf{{p}}^{(gr)} =0$
%for $\alpha =1$;   for $\alpha =2$ the momentum $\mathbf{p}^{(gr)}$ is equal to the coordinate
%of the second cone $ \mathbf{p}_{deg}$ while $p_y^{(in)}$ is the conserving $y$-projection of
%the incident quasi-particle (in the latter case one should change $\mathbf{p} =0$ in the Bolch
%factors in  Eq.(\ref{garphBlochfunctions}) to  $ \mathbf{p}_{deg}$ ).

 Changing the variables
$\zeta-q \rightarrow \zeta$ one gets
\begin{eqnarray}
J_{\alpha}^{(gr)}=e^{-q x/\hbar}\int_{0}^{\infty} \frac{ e^{-x \zeta/ \hbar}}{\varepsilon \mp
i v\sqrt{\zeta (\zeta +2 q}}d\zeta
 \label{AppgraphIntegral2}
\end{eqnarray}

As one sees from Eq.(\ref{AppgraphIntegral2}) the main contribution of the integrand  to the
integral is at $\zeta  \lesssim \hbar/x$.  This inequality means that the square root in the
integral denominator is much less than $\varepsilon /v$ (note that $|p_y^{(in)}\  \lesssim
\varepsilon/v$). Therefore, neglecting the term with the square root one easily takes the
integral and obtains the result presented in Eq.(\ref{GraphPsiFinal}) of the main text in
which constants $B_\alpha$ are
\begin{eqnarray}
B_{\alpha}^{(gr;\pm)}= \nonumber \\
\frac{\hbar^2}{2m}\int_{-\infty}^{\infty}d\bar{y}\Psi^{\prime}_x(0,\bar{y})
\varphi_{\alpha;p_y^{(gr)},p_y^{(in)}}^{(gr;\pm)}(0,\bar{y})d \bar{y}
 \label{graphB1}
\end{eqnarray}

\end{appendix}

\end{document}